\documentclass[aps,preprint,nofootinbib]{revtex4}
\usepackage{amsfonts}
\usepackage{amsmath}
\usepackage{amssymb}
\usepackage{graphicx}%
\providecommand{\U}[1]{\protect\rule{.1in}{.1in}}

\begin{document}
\preprint{ }
\title[ ]{Three-photon annihilation of the electron-positron pairs}
\author{Alexei M. Frolov}
\email{afrolov@uwo.ca}
\affiliation{Department of Chemistry, University of Western Ontario, London, Canada}

\begin{abstract}
Three-photon annihilation of the electron-positron pairs (= $(e^{-}, e^{+}%
)-$pairs) is considered in the electron rest frame. The energy of the incident
positron can be arbitrary. The analytical expression for the cross-section of
three-photon annihilation of the $(e^{-},e^{+})-$pair has been derived and investigated.

\end{abstract}
\volumeyear{ }
\volumenumber{ }
\issuenumber{ }
\eid{ }
\maketitle

In this work we discuss three-photon annihilation of the electron-positron
pairs (or $(e^{-}, e^{+})-$pairs, for short). Annihilation of the $(e^{-},
e^{+})-$pair is considered in the electron rest frame. The non-relativistic
limit for the three-photon annihilation rate $\Gamma_{3 \gamma}$ was obtained
long ago \cite{Lif} - \cite{Nor} (see also \cite{Mac} and \cite{Smi}). In our
earlier work \cite{Fro07} we considered three-photon annihilation of the
electron-positron at arbitrary energies of the colliding particles. However,
due to a number of reasons the closed expression for the three-photon
annihilation cross-section was not derived in \cite{Fro07}. In this study we
want to make the final step in our consideration of the three-photon
annihilation and obtain the closed analytical formula for the cross-section
of the three-photon annihilation. The analysis of this process is performed
in the electron rest frame, while the energy of the incident positron can be
arbitrary.

The principal conservation law in the case of three-photon annihilation of the
free $(e^{-},e^{+})-$pair is
\begin{equation}
p_{1} + p_{2} = k_{1} + k_{2} + k_{3}\label{e1}%
\end{equation}
where $p_{1}$ and $p_{2}$ are the 4-vectors of electron and positron momenta,
respectively. For these two particles we always have $p^{2}_{1} = m^{2}$ and
$p^{2}_{2} = m^{2}$, where $m$ is the electron/positron mass (see, e.g.,
\cite{AB}). In the electron rest frame $p_{1} = (E_{1}, 0) = (m, 0)$. In
Eq.(\ref{e1}) the three four-vectors $k_{1}, k_{2}, k_{3}$ designate the
4-vectors of photon momenta. Here and below $k_{i} = (\omega_{i},
\mathbf{k}_{i})$. For the real photons one finds $k^{2}_{i} = 0$ (so-called
on-shell condition) and $\omega^{2}_{i} = \mathbf{k}^{2}_{i}$ ($i$ = 1, 2, 3).

According to the rules of QED, the corresponding $S-$matrix element ($S_{fi}$)
in the momentum space is
\begin{equation}
S_{fi} = \frac{e^{3}}{V^{3} \sqrt{V}} \sqrt{\frac{m^{2}}{E_{1} E_{2}}}
\sqrt{\frac{(4 \pi)^{3}}{8 \omega_{1} \omega_{2} \omega_{3}}} (2 \pi)^{4}
\delta^{4}(p_{1} + p_{2} - k_{1} - k_{2} - k_{3}) M_{fi}\label{S}%
\end{equation}
where $e$ is the electron charge, while $V$ is the finite normalization
volume, $E_{1}$ and $E_{2}$ are the energies of the incident electron and
positron, respectively. Also, in this equation $\omega_{i}$ ($i$ = 1, 2, 3)
are the three frequencies/energies of the emitted photons. The notation
$M_{fi}$ designates the matrix element which corresponds to the Feynman
diagram of the three-photon annihilation (see, e.g., the diagram presented in
\$ 89 of \cite{AB}). The explicit form of the matrix element $M_{fi}$ is
\begin{equation}
M_{fi} = \imath\overline{v} \Bigl[ \epsilon_{3} \frac{1}{p_{1} - k_{1} - k_{2}
- m} \epsilon_{2} \frac{1}{p_{1} - k_{1} - m} \epsilon_{1} + \epsilon_{2}
\frac{1}{p_{1} - k_{1} - k_{3} - m} \epsilon_{3} \frac{1}{p_{1} - k_{1} - m}
\epsilon_{1} + \ldots\label{M}%
\end{equation}
\begin{equation}
+ \epsilon_{i} \frac{1}{p_{1} - k_{j} - k_{l} - m} \epsilon_{j} \frac{1}{p_{1}
- k_{l} - m} \epsilon_{l} + \ldots\Bigr] u\nonumber
\end{equation}
where $(i, j, l)$ = (1, 2, 3). The $v$ and $u$ are the positron and electron
bi-spinors, respectively, while $k_{i}$ and $\epsilon_{i}$ ($i$ = 1, 2, 3) are
the momentum and polarization of the $i-$th photon. The total number of terms
in the amplitude $M$ equals six. Each of these six terms in Eq.(\ref{M}) can
be transformed in the following way, e.g., for the first term
\begin{equation}
M_{321} = \overline{v} \Bigl[ \epsilon_{3} \frac{1}{p_{1} - k_{1} - k_{2} - m}
\epsilon_{2} \frac{1}{p_{1} - k_{1} - m} \epsilon_{1} \Bigr] u = A_{321}
\overline{v} \Bigl[ \epsilon_{3} (p_{1} + k_{1} + k_{2} + m) \epsilon_{2}
\times\label{M1}
\end{equation}
\begin{equation}
(p_{1} - k_{1} + m) \epsilon_{1} \Bigr] u = A_{321} \overline{v}
\Bigl[ \epsilon_{3} (- p_{2} + k_{3} + m) \epsilon_{2} (p_{1} - k_{1} + m)
\epsilon_{1} \Bigr] u\nonumber
\end{equation}
where
\begin{equation}
A_{321} = \frac{1}{4 (p_{1} \cdot k_{1}) (p_{2} \cdot k_{3})}\label{Norm}%
\end{equation}
Here and below, all matrix elements $M_{ijk}$ and normalization factors
$A_{ijk}$ are designated with the use of three photon indexes which are
uniformly related to the corresponding photon lines on Feynman diagram (by
reading them from the left to the right).

The cross-section of the three-photon annihilation is
\begin{equation}
d\sigma= \int\frac{V^{2} \mid S_{fi} \mid^{2} }{T V \mid\mathbf{v} \mid} V
\frac{d^{3}k_{1}}{(2 \pi)^{3}} V \frac{d^{3}k_{2}}{(2 \pi)^{3}} V \frac
{d^{3}k_{3}}{(2 \pi)^{3}}\label{sigma1}%
\end{equation}
\begin{equation}
= \frac{e^{6}}{(2 \pi)^{5}} \frac{m^{2}}{E_{1} E_{2}} \frac{(4 \pi)^{3}}%
{\mid\mathbf{v} \mid} \int\delta^{4}(p_{1} + p_{2} - k_{1} - k_{2} - k_{3})
\mid M_{fi} \mid^{2} \frac{d^{3}k_{1}}{2 \omega_{1}} \frac{d^{3}k_{2}}{2
\omega_{2}} \frac{d^{3}k_{3}}{2 \omega_{3}}\nonumber
\end{equation}
where in our present case $\mathbf{v} = \frac{\mathbf{p}_{2}}{E_{2}}$ is the
velocity of incoming positron, while $T$ designates a finite (but very large)
time. To finish these calculations one needs to obtain the closed analytical
formulas for the $M_{fi}$ matrix element. In general, such calculations is not
an easy task. However, to simplify these calculations and final expressions
for the amplitudes $M_{321}$ and $M_{ijk}$ one can introduce a few additional
conditions on the photon polarization 4-vectors $\epsilon_{i}$ ($i$ = 1, 2,
3). Below we shall assume that the following conditions are obeyed for the
three polarization 4-vectors of photons $\epsilon_{i}$ \cite{Grein}
\begin{equation}
\epsilon_{i} \cdot k_{i} = 0 \; \; \; , \; \; \; \epsilon_{i} \cdot
\epsilon_{i} = -1 \; \; \; , \; \; \; \epsilon_{i} \cdot p_{1} = 0\label{cond}%
\end{equation}
where $i$ = 1, 2, 3. In this work $p_{1} = (m, 0)$ and we can choose
$\epsilon_{i} = (0, \mathbf{\varepsilon}_{i})$, where $\mathbf{\varepsilon
}_{i}$ are the three unit-norm 3D-vectors.

Now, by applying the conditions, Eq.(\ref{cond}), and using relation $a b = 2
(a \cdot b) - b a$, where $a$ and $b$ are the two arbitrary 4-vectors, one
finds
\begin{equation}
( p_{1} - k_{1} + m ) \epsilon_{1} u = - k_{1} \epsilon_{1} u + \epsilon_{1} (
- p_{1} + m ) u = - k_{1} \epsilon_{1} u
\end{equation}
since $( p_{1} - m ) u = 0$, where $u$ is the electron bi-spinor. Analogously,
since $0 = \overline{v} ( p_{2} + m )$, where $v$ is the positron bispinor, we
can simplify computations of Eq.(\ref{M1}). Finally, for the $M_{321}$ matrix
element one finds
\begin{equation}
M_{321} = - A_{321} \cdot\overline{v} [ \epsilon_{3} k_{3} \epsilon_{2} k_{1}
\epsilon_{1} ] u + 2 A_{321} (\epsilon_{3} \cdot p_{2}) \overline{v} [
\epsilon_{2} k_{1} \epsilon_{1} ] u
\end{equation}
\begin{equation}
= A_{321} \cdot\overline{v} [ 2 (\epsilon_{3} \cdot p_{2}) \epsilon_{2} k_{1}
\epsilon_{1} - \epsilon_{3} k_{3} \epsilon_{2} k_{1} \epsilon_{1}] u\nonumber
\end{equation}
In the case of the $(ijl)-$diagram the analogous expression for the $M_{ijl}$
matrix element takes the form
\begin{equation}
M_{ijl} = A_{ijl} \cdot\overline{v} [ 2 (\epsilon_{i} \cdot p_{2})
\epsilon_{j} k_{l} \epsilon_{l} - \epsilon_{i} k_{i} \epsilon_{j} k_{l}
\epsilon_{l}] u
\end{equation}
where $i \ne j \ne l$ = (1,2,3) and
\begin{equation}
A_{ijl} = \frac{1}{4 (p_{1} \cdot k_{l}) (p_{2} \cdot k_{i})}\label{Norm1}%
\end{equation}
The conjugate amplitude $M^{*}_{ijl}$ is
\begin{equation}
M^{*}_{ijl} = A_{ijl} \cdot\overline{u} [ 2 (\epsilon_{i} \cdot p_{2})
\epsilon_{l} k_{l} \epsilon_{j} - \epsilon_{l} k_{l} \epsilon_{j} k_{i}
\epsilon_{i}] v
\end{equation}

The expression for the $\mid M \mid^{2}$ value is reduced to the sum of the
six matrix element $M^{*}_{ijl} M_{321}$, where $(i, j, l)$ = (1, 2, 3). The
analytical formula for the $M^{*}_{ijl} M_{321}$ matrix element averaged over
the initial electron and positron states can be written in the form
\begin{equation}
M^{*}_{ijl} M_{321} = A_{ijl} A_{321} \cdot B_{ijl} = \frac{A_{ijl}
A_{321}}{16 m^{2}} ( B_{1} - m^{2} B_{2} )
\end{equation}
where
\begin{equation}
B_{1} = 4 (\epsilon_{i} \cdot p_{2}) (\epsilon_{3} \cdot p_{2}) B_{1a} - 2
(\epsilon_{i} \cdot p_{2}) B_{1b} - 2 (\epsilon_{3} \cdot p_{2}) B_{1c} +
B_{1d}%
\end{equation}
and
\begin{equation}
B_{2} = 4 (\epsilon_{i} \cdot p_{2}) (\epsilon_{3} \cdot p_{2}) B_{2a} - 2
(\epsilon_{i} \cdot p_{2}) B_{2b} - 2 (\epsilon_{3} \cdot p_{2}) B_{2c} +
B_{2d}%
\end{equation}
where the explicit expressions for the eight traces $B_{1k}$ and $B_{2k}$ ($k
= a, b, c, d$) are
\begin{equation}
B_{1a} = Tr[p_{2} \epsilon_{l} k_{l} \epsilon_{j} p_{1} \epsilon_{2} k_{1}
\epsilon_{1}] \; \; \; \; \; \; \; \; \; , \; \; \; \; \; \; \; \; B_{2a} =
Tr[\epsilon_{l} k_{l} \epsilon_{j} \epsilon_{2} k_{1} \epsilon_{1}]\label{e32}%
\end{equation}
\begin{equation}
B_{1b} = Tr[p_{2} \epsilon_{l} k_{l} \epsilon_{j} p_{1} \epsilon_{3} k_{3}
\epsilon_{2} k_{1} \epsilon_{1}] \; \; \; \; \; \; , \; \; \; \; \; B_{2b} =
Tr[\epsilon_{l} k_{l} \epsilon_{j} \epsilon_{3} k_{3} \epsilon_{2} k_{1}
\epsilon_{1}]
\end{equation}
\begin{equation}
B_{1c} = Tr[p_{2} \epsilon_{l} k_{l} \epsilon_{j} k_{i} \epsilon_{i} p_{1}
\epsilon_{2} k_{1} \epsilon_{1}] \; \; \; \; \; \; , \; \; \; \; \; \; B_{2c}
= Tr[\epsilon_{l} k_{l} \epsilon_{j} k_{i} \epsilon_{i} \epsilon_{2} k_{1}
\epsilon_{1}]
\end{equation}
\begin{equation}
B_{1d} = Tr[p_{2} \epsilon_{l} k_{l} \epsilon_{j} k_{i} \epsilon_{i} p_{1}
\epsilon_{3} k_{3} \epsilon_{2} k_{1} \epsilon_{1}] \; \; \; , \; \; \; B_{2d}
= Tr[\epsilon_{l} k_{l} \epsilon_{j} k_{i} \epsilon_{i} \epsilon_{3} k_{3}
\epsilon_{2} k_{1} \epsilon_{1}]\label{e39}%
\end{equation}
Thus, the problem of three-photon annihilation of the electron-positron pair
at arbitrary energies of the colliding particles is reduced to the analytical
computation of these eight traces. The explicit formulas for all individual
traces, Eq.(\ref{e32}) - Eq.(\ref{e39}), as well as for $B_{ijl}$ can be
obtained directly from the author (some restrictions may apply).

The formulas given above correspond to the case when the polarizations of all
photons (i.e. $\epsilon_{i}$, $i$ = 1, 2, 3) are known. If this is not the
case, then in the formulas presented above one needs to compute the sums over
polarizations of all final photons. Below, we perform the polarization
summation using the standard replacement $\sum_{\lambda=0,3} \epsilon_{\mu
}^{(\lambda)} \epsilon_{\nu}^{(\lambda)} = -g_{\mu\nu}$. However, the last
condition in Eq.(\ref{cond}) ($\epsilon_{i} \cdot p_{1} = 0$) which can
contradict such a replacement. Following the procedure developed in
\cite{Grein} one finds that there is no contradiction in the electron rest
frame. This means that the standard replacement $\sum_{\lambda=0,3}
\epsilon_{\mu}^{(\lambda)} \epsilon_{\nu}^{(\lambda)} = -g_{\mu\nu}$ can be
used in this study to perform the polarization summation.

After computing all traces summed over photon polarizations, the explicit
expression for the $M^{*}_{ijl} M_{321}$ value is written in the form
\begin{equation}
M^{*}_{ijl} M_{321} = \frac{A_{ijl} \cdot A_{321}}{16 m^2} D_{ijl}
\end{equation}
for $(i,j,l)$ = (1,2,3) and $D_{ijl}$ denotes the resulting traces computed
for each of these cases. The explicit formulas for $D_{ijl}$ are
\begin{equation}
D_{123} = 32 m^2 [ 2 (k_{1}\cdot k_{3})^{2} + (k_{1} \cdot p_{2}) (k_{3}
\cdot p_{1}) + (k_{1} \cdot p_{1}) (k_{3} \cdot p_{2}) + (k_{1} \cdot
k_{3}) (2 m^{2} - p_{1} \cdot p_{2}) ]\label{eqd1}%
\end{equation}
\begin{equation}
D_{132} = 32 m^2 [ (k_{1} \cdot k_{3}) (k_{2} \cdot p_{1}) - (k_{1} \cdot
p_{2}) (4 k_{2} \cdot k_{3} + k_{2} \cdot p_{1} - 2 k_{2} \cdot p_{2}) +
(k_{1} \cdot p_{1}) (- k_{2} \cdot k_{3}\label{eqd2}%
\end{equation}
\begin{equation}
+ k_{2} \cdot p_{2}) + (k_{1} \cdot k_{2}) (k_{3} \cdot p_{1}) + (k_{1} \cdot
k_{2}) (p_{1} \cdot p_{2}) ]\nonumber
\end{equation}
\begin{equation}
D_{213} = 32 m^2 [ - (k_{1} \cdot k_{2}) (k_{3} \cdot p_{1}) + (k_{1} \cdot
p_{2}) (k_{3} \cdot p_{1}) + (k_{1} \cdot p_{1}) (k_{2} \cdot k_{3} - k_{3}
\cdot p_{2})\label{eqd3}
\end{equation}
\begin{equation}
- 4 (k_{1} \cdot k_{2}) (k_{3} \cdot p_{2}) + 2 (k_{1} \cdot p_{2}) (k_{3}
\cdot p_{2}) + (k_{1} \cdot k_{3}) (k_{2} \cdot p_{1} + p_{1} \cdot p_{2})
]\nonumber
\end{equation}
\begin{equation}
D_{231} = 64 (k_{1} \cdot p_{2}) [ (k_{1} \cdot p_{1}) (- m^{2}
 - 2 k_{2} \cdot k_{3} + k_{2} \cdot p_{2} + k_{3} \cdot p_{2}) - (k_{1}
 \cdot k_{2} + k_{1} \cdot k_{3}) (p_{1} \cdot p_{2})\label{eqd4}%
\end{equation}
\begin{equation}
 + (k_{1} \cdot p_{2}) (k_{2} \cdot p_{1} + k_{3} \cdot p_{1} + 2 p_{1} \cdot
 p_{2}) ]\nonumber
\end{equation}
\begin{equation}
 D_{312} = 64 (k_{1} \cdot k_{2}) [ m^{4}+2 k_{3} \cdot p_{1}
 (m^{2} + k_{3} \cdot p_{2}) + 2 m^{2} (p_{1} \cdot p_{2}) - (k_{3} \cdot
 p_{2}) (m^{2} + 4 p_{1} \cdot p_{2}) ]\label{eqd5}%
\end{equation}
\begin{equation}
 D_{321} = - 128 (k_{1} \cdot p_{2}) [ (k_{1} \cdot p_{2}) (k_{3}
 \cdot p_{1}) + (k_{1} \cdot p_{1}) (m^{2} - k_{3} \cdot p_{2}) - (k_{1} \cdot
 k_{3}) (k_{3} \cdot p_{1} + p_{1} \cdot p_{2}) ]\label{eqd6}%
\end{equation}
In actual computations the sum $(M_{123} + M_{132} + M_{213} + M_{231} +
M_{312} + M_{321}) \cdot M_{321}$ must be multiplied by an additional
factor (= degeneracy factor) $S = \frac{1}{3!} = \frac{1}{6}$. This factor
is needed, since the final state in the case of three-photon annihilation
contains three identical photons. This means that the $\mid M_{fi} \mid^2$
term from Eq.(\ref{e1}) is represented in the form
\begin{equation}
 \mid M_{fi} \mid^2 = \frac{1}{6} (M_{123} + M_{132} + M_{213} + M_{231} +
 M_{312} + M_{321}) \cdot M_{321}
\end{equation}

Now, we want to obtain the explicit formula for the cross-section of the
three-photon annihilation. First, consider the computation of the following
auxiliary integral
\begin{equation}
I = \int\frac{d^{3}k_{1}}{2 \omega_{1}} \frac{d^{3}k_{2}}{2 \omega_{2}}
\frac{d^{3}k_{3}}{2 \omega_{3}} \delta^{4}(p_{1} + p_{2} - k_{1} - k_{2} -
k_{3}) \cdot f(k_{1}, k_{2}, k_{3})\label{ai}%
\end{equation}
where $f(x,y,z)$ is an arbitrary, in principle, function of three variables
$x, y$ and $z$. The 3D-integral over the $k_{3}$ is reduced to the following
four-dimensional integral with the use of the relation (see, e.g., \cite{AB})
\begin{equation}
\frac{d^{3}k_{3}}{2 \omega_{3}} = \int_{-\infty}^{+\infty} d^{4}k_{3}
\delta(k_{3} \cdot k_{3} - 0) \Theta[(k_{3})_{0}]
\end{equation}
where $\Theta(x) = 1$, if $x \ge0$ and zero otherwise. The notation $(b)_{0}$
means 0-component of the four-vector $b$. With the use of this formula the
expression, Eq.(\ref{ai}), takes the form
\begin{equation}
dI = \int\frac{d^{3}k_{1}}{2 \omega_{1}} \frac{d^{3}k_{2}}{2 \omega_{2}}
\delta[(p_{1} + p_{2} - k_{1} - k_{2})^{2}] \Theta(E_{1} + E_{2} -
\omega_{1} - \omega_{2}) \cdot f(k_{1}, k_{2}, p_{1} + p_{2} - k_{1} -
k_{2})\nonumber
\end{equation}
\begin{equation}
= \frac{1}{4} \int \omega_{1} d\omega_{1} d\Omega_{1} d\Omega_{2}
\int_{0}^{a} \omega_{2} d\omega_{2} f(k_{1}, k_{2}, p_{1} + p_{2} - k_{1}
- k_{2}) \delta[(p_{1} + p_{2} - k_{1} - k_{2})^{2}]\label{ai1}%
\end{equation}
\begin{equation}
= \frac{1}{8} \int \omega_{1} d\omega_{1} d\Omega_{1} d\Omega_{2}
\int_{0}^{a} \omega_{2} d\omega_{2} f(k_{1}, k_{2}, p_{1} + p_{2} - k_{1}
- k_{2}) \delta[m^{2} + E_{1} E_{2} - \mathbf{p}_{1}
\cdot\mathbf{p}_{2}\nonumber
\end{equation}
\begin{equation}
- (E_{1} + E_{2}) (\omega_{1} + \omega_{2}) + \omega_{1} \omega_{2} -
\mathbf{k}_{1} \cdot\mathbf{k}_{2} + (\mathbf{k}_{1} + \mathbf{k}_{2})
\cdot(\mathbf{p}_{1} + \mathbf{p}_{2}) ]\nonumber
\end{equation}
where $a = E_{1} + E_{2} - \omega_{1} (\geq0)$. The total number of
variables in this (final) formula is five (5 = 9 - 4) as expected.

In our present case $p_{1}=(m,0)$, i.e. $E_{1}=m$ and such a substitution
simplifies all following formulas. In particular, for the auxiliary integral
$I$ one finds
\begin{equation}
 dI = \frac{1}{8}\omega_{1}d\omega_{1}d\Omega_{1}d\Omega_{2}\cdot F\cdot
\frac{m^{2}+mE_{2}-(m+E_{2})\omega_{1}+\mid\mathbf{p}_{2}\mid\omega_{1}%
\cos\Theta_{1}}{(m+E_{2}+\mid\mathbf{p}_{2}\mid\cos\Theta_{2}+\omega
_{1}-\omega_{1}\cos{\Psi}_{12})^{2}}\label{ai2}%
\end{equation}%
\begin{equation}
=\frac{1}{8} F \cdot\frac{m^{2} + m E_{2} - (m + E_{2}) \omega_{1}
+\mid\mathbf{p}_{2}\mid\omega_{1}\cos\Theta_{1}}{(m + E_{2} +
 \mid\mathbf{p}_{2}\mid \cos\Theta_{2} + \omega_{1} - \omega_{1}
 \cos{\Psi}_{12})^{2}} \omega_{1} d\omega_{1} sin\Theta_{1} d\Theta_{1}
 d\phi_{1} sin\Theta_{2} d\Theta_{2} d\phi_{2} \nonumber
\end{equation}
where $\mid\mathbf{p}_{2}\mid=\sqrt{E_{2}^{2}-m^{2}}$ and notation $F$ stands
for the function $f$ from Eq.(\ref{ai1}) which also contains all mentioned
substitutions for the $k_{3}$ and $\omega_{2}$ variables. The notations
$\Theta_{1}$ and $\Theta_{2}$ mean the two angles between the positron and
first and second photon, respectively. The notation $\Psi_{12}$ designates the
angle between the two photons (photons 1 and 2). Computation of the $cos{\Psi
}_{12}$ value is performed with the use of addition theorem for spherical
harmonics (see, e.g., \cite{Gel})
\begin{equation}
 cos{\Psi}_{12} = cos\Theta_{1} cos\Theta_{2} + sin\Theta_{1} sin\Theta_{2}
 cos(\phi_{1}-\phi_{2})
\end{equation}

\[
=cos\Theta_{1}cos\Theta_{2}+sin\Theta_{1}sin\Theta_{2}cos\phi_{1}cos\phi
_{2}-sin\Theta_{1}sin\Theta_{2}sin\phi_{1}sin\phi_{2}%
\]
The following integration over four angular variables does not present any
difficulty.

Now, by using the explicit formulas Eq.(\ref{Norm}), Eq.(\ref{sigma1}) and
Eq.(\ref{eqd1}) - Eq.(\ref{eqd6}) one can finish these calculations and
obtain the final formulas for the cross-section of the three-photon
annihilation. The complete formula which contains all terms is extremely
complicated. In fact, its derivation has been made with the use a symbolic
algebra system, such as Maple \cite{Mapl}. Here we want to illustrate our
calculations only for one term in $\mid M \mid^{2}$. Calculations of other
five terms are almost identical. Let us present the final formula for the
differential cross-section of the three-photon annihilation. In the electron
rest frame one finds from Eq.(\ref{sigma1}) and Eq.(\ref{ai2})
\begin{equation}
\frac{d\sigma}{d\omega_{1} d\Omega_{1} d\Omega_{2}} = \frac{e^{6}}{(2 \pi
)^{5}} \frac{m \omega_{1}}{E_{2}} \frac{(4 \pi)^{3}}{\mid\mathbf{v} \mid} \mid
M \mid^{2} \cdot\frac{m^{2} + m E_{2} - (m + E_{2}) \omega_{1} +
\mid\mathbf{p}_{2} \mid\omega_{1} \cos\Theta_{1}}{(m + E_{2} + \mid
\mathbf{p}_{2} \mid\cos\Theta_{2} + \omega_{1} - \omega_{1} \cos{\Psi}%
_{12})^{2}}\label{fin1}%
\end{equation}
where $\mid M \mid^{2} = \frac16 (M^{*}_{123} + M^{*}_{132} + M^{*}_{213} +
M^{*}_{231} + M^{*}_{312} + M^{*}_{321}) \cdot M_{321}$. For simplicity,
consider only the first term $M^{*}_{123} \cdot M_{321}$ (analysis of other
terms is very similar). For the $M^{*}_{123} \cdot M_{321}$ term one finds
from formulas presented above
\begin{equation}
M^{*}_{123} M_{321} = A_{123} \cdot A_{321} D_{123} = \frac{8 Q}{R}
\end{equation}
where
\begin{equation}
Q = 2 [k_{1} \cdot(p_{1} + p_{2} - k_{1} - k_{2})]^{2} + (k_{1} \cdot p_{2})
[p_{1} \cdot(p_{1} + p_{2} - k_{1} - k_{2})] + (k_{1} \cdot p_{1}) [p_{2}
\cdot(p_{1}
\end{equation}
\begin{equation}
+ p_{2} - k_{1} - k_{2})] + (2 m^{2} - p_{1} \cdot p_{2}) ([k_{1} \cdot(p_{1}
+ p_{2} - k_{1} - k_{2})]\nonumber
\end{equation}
\begin{equation}
= 2 \omega^{2}_{1} [m + E_{2} - \mid\mathbf{p}_{2} \mid cos\Theta_{1} -
\omega_{2} + \omega_{2} \cos{\Psi}_{12}]^{2} + m \omega_{1} (m + E_{2} -
\omega_{1} - \omega_{2}) (E_{2} - \mid\mathbf{p}_{2} \mid cos\Theta
_{2})\nonumber
\end{equation}
\begin{equation}
+ m \omega_{1} [m^{2} + E_{2} m + \mid\mathbf{p}_{2} \mid\omega_{1}
cos\Theta_{1} - \mid\mathbf{p}_{2} \mid\omega_{2} cos\Theta_{2} - E_{2}
(\omega_{1} + \omega_{2})] + m \omega_{1} (2 m - E_{2}) (m + E_{2}\nonumber
\end{equation}
\begin{equation}
- \mid\mathbf{p}_{2} \mid\omega_{2} cos\Theta_{1} + \omega_{2} \cos{\Psi}%
_{12})\nonumber
\end{equation}
and
\begin{equation}
R = [p_{1} \cdot(p_{1} + p_{2} - k_{1} - k_{2})] (p_{2} \cdot k_{1}) (p_{1}
\cdot k_{1}) [p_{2} \cdot(p_{1} + p_{2} - k_{1} - k_{2})]
\end{equation}
\begin{equation}
= m^{2} E_{2} \omega^{2}_{1} (m + E_{2} - \omega_{1} - \omega_{2}) (m^{2} + m
E_{2} - E_{2} \omega_{1} - E_{2} \omega_{2} + \mid\mathbf{p}_{2} \mid
\omega_{1} cos\Theta_{1} + \mid\mathbf{p}_{2} \mid\omega_{2} \cos\Theta
_{2})\nonumber
\end{equation}
where the function $\omega_{2}$ is
\begin{equation}
\omega_{2} = \frac{m^{2} + m E_{2} - (m + E_{2}) \omega_{1} + \mid
\mathbf{p}_{2} \mid\omega_{1} cos\Theta_{1}}{m + E_{2} + \mid\mathbf{p}_{2}
\mid cos\Theta_{2} + \omega_{1} - \omega_{1} \cos{\Psi}_{12}}%
\end{equation}
and $\mid\mathbf{p}_{2} \mid= \sqrt{E^{2}_{2} - m^{2}}$. Note that all these
formulas contain only five variables already mentioned above $\omega_{1},
cos\Theta_{1}, cos\Theta_{2}$ and $cos\Psi_{12}$ (i.e. $\phi_{1}$ and
$\phi_{2}$). The same conclusion is true about the five other terms in the
formula for the $\mid M \mid^{2}$ factor in Eq.(\ref{fin1}). Formally, after
the integration over these five variables ($\omega_{1}, cos\Theta_{1},
cos\Theta_{2}, \phi_{1}$ and $\phi_{2}$) the final expression (total
cross-section of the three-photon annihilation of the ($e^{-},e^{+}$)-pair) is
the function of only one actual variable $\gamma_{p} = \frac{E_{2}}{m}$, where
$\gamma_{p}$ is the Lorentz gamma-factor of the incoming (= fast) positron.
The well known Dirac's formula for the total cross-section of the two-photon
annihilation of the ($e^{-},e^{+}$)-pair also depends upon the positron
gamma-factor $\gamma_{p}$ only, if it is written in the electron rest frame
\cite{Dir}.

The formulas given above allow one to describe and determine the angular and
energy distribution of the photons emitted during three-photon annihilation of
the electron-positron pair (= $(e^{-},e^{+})$-pair). Our results are obtained
in the electron rest frame. The energy of the incident positron $E_{2} (\equiv
E_{p})$ can be arbitrary. The analytical expression has finally been derived
for the differential cross-section of three-photon annihilation of the
electron-positron pair. It is shown that such a cross-section is an explicit
function of the five variables ($\omega_{1}, cos\Theta_{1}, cos\Theta_{2},
\phi_{1}$ and $\phi_{2}$) which describe three-photon annihilation of the
$(e^{-},e^{+})-$pair. In fact, our formula for the differential cross-section
of three-photon annihilation of the electron-positron pair can be simplified
even further, since it contains only four actual variables ($\omega_{1},
cos\Theta_{1}, cos\Theta_{2}$ and $\phi_{1} - \phi_{2}$). In our next study we
want to consider the four-photon annihilation of the $(e^{-},e^{+})-$pair.

\subsection{The non-relativistic limit of the differential cross-section}

Let us obtain the explicit formula for the differential cross-section
$d\sigma_{3\gamma}$ at non-relativistic energies of the colliding particles
(or positron in our case). First, note that in the non-relativistic limit
one finds for the $\omega_2$ frequency
\begin{equation}
  \omega_2 = \frac{m^2 + m E_2 - (m + E_2) \omega_1}{(m + E_2 +
  \omega_1 - \omega_1 \cos {\Psi}_{12})^2} =
  \frac{(m + E_2) (m - \omega_1)}{m + E_2 + \omega_1 - \omega_1
 \cos {\Psi}_{12}}
\end{equation}
In other words, in the non-relativistic limit the $\omega_2$ frequency does
not depend any photon-positron angles $\Theta_1, \phi_1$ and/or $\Theta_2,
\phi_2$. However, it explicitly depends upon the inter-electron angle
$\cos {{\Psi}_{12}}$.

The final formula for the differential cross-section of the three-photon
annihilation is written in the form (in the electron rest frame)
\begin{eqnarray}
  \frac{d\sigma_{3 \gamma}}{d\omega_1 d\Omega_1 d\Omega_2} = \frac{2
 e^6}{\pi^2} \frac{\omega_1}{m^2 \gamma_p \mid {\bf v} \mid} \mid M \mid^2
 \cdot \frac{m^2 + m E_2 - (m + E_2) \omega_1 + \mid {\bf p}_2 \mid \omega_1
 \cos \Theta_1}{(m + E_2 + \mid {\bf p}_2 \mid \cos \Theta_2 + \omega_1 -
 \omega_1 \cos {\Psi}_{12})^2} \label{fin1} \\
 = \frac{2 e^6}{6 \cdot 8 \pi^2 m^2 \gamma_p \mid {\bf v} \mid}
 \mid (M_{123} + M_{132} + M_{213} + M_{231} + M_{312} + M_{321}) \cdot
 M_{321} \mid \frac{d \omega_2}{d (cos\Psi_{12})} \nonumber
\end{eqnarray}
where $\mid M \mid^2 = (M^{*}_{123} + M^{*}_{132} + M^{*}_{213} +
M^{*}_{231} + M^{*}_{312} + M^{*}_{321}) \cdot M_{321}$ (the factor
$\frac16$ has been moved in front of Eq.(\ref{fin1})). Now, consider the
non-relativistic limit of the $d\sigma_{3 \gamma}$ cross-section. In the
non-relativistic limit we can integrate over all angular variables, but one
($\Psi_{12}$). This gives us an additional factor $8 \pi^2$ and the formula,
Eq.(\ref{fin1}) takes the form
\begin{equation}
 d\sigma_{3 \gamma} = \frac{2 e^6}{6 m^2 \mid {\bf v}
 \mid} \mid M \mid^2 d\omega_1 \frac{d \omega_2}{d (cos\Psi_{12})}
 d(\cos \Psi_{12}) = \frac{2 e^6}{6 m^2 \mid {\bf v} \mid} \mid M \mid^2
 d\omega_1 d \omega_2 \label{expf}
\end{equation}
where the expression $\omega_3 = 2 m - \omega_1 - \omega_2$ must be used
everywhere in the $\mid M \mid^2$ factor. The formula Eq.(\ref{expf})
essentially coincides with the expression (89.14) from \cite{AB}. Such a
coincidence will be exact, if we can extract an additional factor $4$ from
$\mid M \mid^2$.

The final step in computations of the non-relativistic limit for the
cross-section $d\sigma_{3 \gamma}$ is to obtain the explicit formula for
the $\mid M \mid^2 = (M_{123} + M_{132} + M_{213} + M_{231} + M_{312} +
M_{321}) \cdot M_{321}$. From the non-relativistic relations $k_i + k_j = 2
m - k_l$ one finds that $k_i \cdot k_j = (2 m - k_l)^2$, where $(i, j, l) =
(1, 2, 3)$. Now, we can find the analytical expression for the $m^2 (M_{123}
+ M_{132} + M_{213} + M_{231} + M_{312} + M_{321}) \cdot M_{321}$. For
instance, the formula for the first term (i.e. for the $m^2 M_{123} \cdot
M_{321}$ term) is
\begin{eqnarray}
 m^2 M_{123} \cdot M_{321} = \frac{1}{256 m^4 \omega^{2}_1 \omega^{2}_3}
 \cdot 64 m^4 \Bigl[ 4 (m - \omega_2)^2 + \omega_1 \omega_3 + m (m -
 \omega_2) \Bigr] \\
 = \frac{4 (m - \omega_2)^2 + \omega_1 \omega_3 +
 m (m - \omega_2)}{4 \omega^{2}_1 \omega^{2}_3} \nonumber
\end{eqnarray}
The derivation of explicit formulas for other six matrix elements $m^2
M_{ijl} \cdot M_{321}$ is also straightforward. In turn, the known formulas
for these matrix elements allows one to obtain the non-relativistic limit
of the differential cross-section of the three-photon annihilation of the
$(e^{-},e^{+})-$pair.

\textbf{Acknowledgement}

I would like to thank N. Kiriushcheva and S.V. Kuzmin for useful discussions
and help with the manuscript.


\begin{thebibliography}{99}                                                                                               %
\bibitem {Lif}E.M. Lifshitz, DAN SSSR \textbf{60}, 211 (1948).

\bibitem {Sok}D. Ivanenko and A. Sokolov, DAN SSSR \textbf{61}, 51 (1948).

\bibitem {Nor}A. Ore and J.L. Powel, Phys. Rev. \textbf{75}, 1696 (1949).

\bibitem {Mac}W.H. McMaster, Rev. Mod. Phys. \textbf{33}, 8 (1961).

\bibitem {Smi}A.I. Smirnov, Russian Phys. Journal \textbf{18}, 709 (1975).

\bibitem {Fro07}A.M. Frolov and F.A. Chishtie, J. Phys. A \textbf{74}, 11923 (2007).

\bibitem {AB}V.B. Berestetskii, E.M. Lifshitz and L.P. Pitaevskii, Quantum
Electrodynamics, (2nd ed., Pergamon Press, New York, (1982)).

\bibitem {Grein}W. Greiner and J. Reinhardt, \textit{Quantum Electrodynamics},
(Springer, Berlin, 3rd Ed., (2003)).

\bibitem {Gel}I.M. Gelfand, R.A. Minlos and Z.Ya. Shapiro,
\textit{Representations of the Rotation and Lorentz Groups and Their
Applications}, (McGraw-Hill Book Company, New York (1963)).

\bibitem {Mapl}A product of Waterloo Maple Inc., Waterloo, Ontario, Canada
(see: http://www.maplesoft.com).

\bibitem {Dir}P.A.M. Dirac, Proc. Cambr. Phil. Soc. \textbf{26}, 361 (1930).
\end{thebibliography}
\end{document}